\documentclass[useAMS,usenatbib]{mn2e}

\usepackage{amssymb}
\usepackage{times}
\usepackage{graphicx}
\usepackage{epstopdf}
\usepackage{subfigure}

\newcommand{\beq}{\begin{equation}}
\newcommand{\eeq}{\end{equation}}

\newcommand{\bfd}{\mathbf{d}}

\newcommand{\bfP}{\mathbf{P}}

\def\gs{\mathrel{\lower0.6ex\hbox{$\buildrel {\textstyle >}\over{\scriptstyle \sim}$}}}
\def\ls{\mathrel{\lower0.6ex\hbox{$\buildrel {\textstyle <}\over{\scriptstyle \sim}$}}}
\newcommand{\simgt}{\lower.5ex\hbox{$\; \buildrel > \over \sim \;$}}
\newcommand{\simlt}{\lower.5ex\hbox{$\; \buildrel < \over \sim \;$}}

\newcommand{\aap}{A\&A}
\newcommand{\apj}{ApJ}
\newcommand{\apjl}{ApJ}

\newcommand{\aj}{AJ}
\newcommand{\pasj}{PASJ}

\newcommand{\mnras}{MNRAS}

\begin{document}

\title[The $c(M)$ at $z\sim 1$]{The mass-concentration relation in massive galaxy clusters at redshift $\sim 1$}
\author[M. Sereno and G. Covone]{
Mauro Sereno$^{1,2}$\thanks{E-mail: mauro.sereno@polito.it (MS)}, Giovanni Covone$^{3,4}$
\\
$^1$Dipartimento di Scienza Applicata e Tecnologia, Politecnico di Torino, corso Duca degli Abruzzi 24, I-10129 Torino, Italia\\
$^2$INFN, Sezione di Torino, via Pietro Giuria 1, I-10125, Torino, Italia\\
$^3$Dipartimento di Scienze Fisiche, Universit\`a di Napoli ``Federico II", Complesso Universitario di Monte S. Angelo, via Cinthia, 80126 Napoli, Italia\\
$^4$INFN, Sezione di Napoli, Complesso Universitario di Monte S. Angelo, via Cinthia, 80126 Napoli, Italia\\
}


\maketitle

\begin{abstract}
Mass and concentration of clusters of galaxies are related and evolving with redshift. We study the properties of a sample of 31 massive galaxy clusters at high redshift, $0.8\ls z \ls 1.5$, using weak and strong lensing observations. Concentration is a steep function of mass, $c_{200} \propto M_{200}^{-0.83 \pm0.39}$, with higher-redshift clusters being less concentrated. Mass and concentration from the stacked analysis, $M_{200}=(4.1\pm0.4)\times 10^{14}M_\odot/h$ and $c_{200}=2.3\pm0.2$, are in line with theoretical results extrapolated from the local universe. Clusters with signs of dynamical activity preferentially feature high concentrations. We discuss the possibility that the whole sample is a mix of two different kinds of haloes. Over-concentrated clusters might be accreting haloes out of equilibrium in a transient phase of compression, whereas less concentrated ones might be more relaxed.
\end{abstract}


\section{Introduction}

Cluster of galaxies are regarded as fundamental cosmological probes and unique astrophysics laboratory. Their successful use relies on accurate measurements of mass and concentration. The hierarchical cold dark matter model with a cosmological constant ($\Lambda$CDM) can explain many features of galaxy clusters. The Navarro-Frenk-White (NFW) density profile \citep{nfw96,nav+al97} reproduces accurately the density profile over most radii, but the actual mass-concentration relation, $c(M)$, is still debated \citep{co+na07,bro+al08}.

The concentration measures the halo central density relative to outer regions and should be related to its virial mass and redshift \citep{bul+al01}. It was early found that lower mass and lower redshift clusters show higher concentrations \citep{bul+al01,duf+al08}. If these theoretical trends were still present at higher mass and redshift, we should not observe the significant number of over-concentrated clusters we actually detect \citep{bro+al08,ogu+al09}. 

Recent high-volume simulations showed somewhat different trends, with a turn-around and increase in concentrations towards very high mass and high redshifts \citep{kly+al11,pra+al11}. However, drawing firm conclusions from these simulations is problematic because some features in the estimated $c(M)$ relation might be an artefact of procedures used to measure the concentrations and of the selection criteria used to bin the haloes \citep{me+ra13}.

Theoretical results are mainly based on dark matter only simulations. Baryonic physics is expected to play a fundamental role in shaping the cluster density profile in the inner regions. However, the explicit effects of cooling, feedback, and baryon-DM interplay on the density profile are still ambiguous \citep{gne+al04,roz+al08,duf+al10,deb+al13}. Cooling makes haloes more concentrated but the effect is mass dependent and strongly affected by the efficiency of feedback processes.

The theoretical $c(M)$ relation depends on the cosmological parameters too \citep{kwa+al12,deb+al13}. Other than the normalisation of the power spectrum and the dark matter content, the dark energy equation of state parameter has an effect too, mainly in lower-mass haloes  \citep{kwa+al12}.

The accurate observational measurement of the $c(M)$ relation can then give important insights on the main processes taking place in galaxy clusters and help to constrain the cosmological parameters. The measured normalisation factor of the $c(M)$ relation of galaxy clusters is usually higher than expected, whereas the slope is steeper \citep{co+na07,ett+al10}. Furthermore, concentrations in lensing clusters are systematically larger than in X-ray luminous ones \citep{co+na07}. 

Here we focus on lensing clusters. Prominent strong lensing clusters are usually found to have very high concentrations \citep{bro+al08,ogu+al09,ume+al11}. This over-concentration problem is partially explained by orientation and shape biases \citep{ogu+al05,se+um11,ras+al12}. Clusters selected according to their gravitational lensing features or X-ray flux may form biased samples preferentially elongated along the line of sight \citep{hen+al07,men+al11}. Triaxial haloes are more powerful lenses than their spherical counterparts \citep{og+bl09} and the strongest lenses are expected to be a highly biased population of haloes preferentially oriented towards the observer. The projected matter density of clusters elongated along the line of sight is relatively higher, and the lensing properties are boosted \citep{hen+al07}. Neglecting halo triaxiality can lead to over- and under-estimates of up to 50 per cent and a factor of 2 in halo mass and concentration, respectively \citep{cor+al09}. The opposite takes place for clusters elongated in the plane of the sky. 

The determination of the shape and orientation of clusters requires deep multi wave-length observations, ranging from X-ray measurements of temperature and brightness to optical observations of lensing and galaxy distribution to measurements of the Sunyaev-Zel'dovich (SZ) effect in the radio \citep{ser+al12a,ser+al13,lim+al12}. However, the orientation bias can not fully account for the discrepancy between theory and observations. \citet{ser+al10} investigated a sample of 10 massive strong lensing (SL) clusters. They found that nearly one half of the clusters was made of low concentration, mildly triaxial lensing clusters with a strong orientation bias, whereas the second half was very concentrated.

Another source of concern is the choice and size of the sample. The over-concentration problem is reduced in larger and better defined samples, but some area of disagreement still persists. \citet{se+zi12} performed a triaxial strong-lensing analysis of the 12 MACS clusters at $z > 0.5$, a complete X-ray selected sample, finding no evidence for over-concentrated clusters. Their analysis suggested that some high-$z$, unrelaxed clusters with noticeable substructures may form a class of prominent strong gravitational lenses with low concentration and relatively shallow inner profiles \citep{zit+al10,zit+al11}. \citet{oka+al10} found that 19 X-ray selected clusters with significant weak lensing signal showed a strong correlation in the $c(M)$ relation with more massive clusters having on average smaller concentrations, although the slope was steeper than theoretical predictcannotions. \citet{ogu+al12} performed a combined strong and weak lensing analysis of 28 clusters from the Sloan Giant Arcs Survey. They found that the observationally inferred concentration is a steep function of the mass and appear to be much higher at lower masses than predictions. 

Some discrepancies between theory and observations are mitigated when stacking techniques are employed to derive the mean properties of clusters. Weak lensing analyses of stacked clusters of lesser mass did not show the high concentration problem \citep{joh+al07,man+al08}. \citet{oka+al13} performed a stacked weak-lensing analysis of a complete and volume-limited sample of 50 massive galaxy clusters at $0.15 \ls z \ls 0.3$. They found shallow density profiles that are broadly in line with predictions from numerical simulations. \citet{ogu+al12} found that the concentration parameter measured with stacked data was slightly smaller than what expected from the $c(M)$ relation constrained with the individual clusters in their sample. They argued that the smaller stacked concentration might be due to the wide range of cluster properties they were averaging on.

In this paper we analyse mass and concentration in a sample of lensing galaxy clusters at $z\sim1$. At this redshift any upturn in the $c(M)$ relation might be noticeable in the massive end of the cluster population, which compensates for the lower lensing strength of clusters at high $z$. The paper is organised as follows. We introduce the sample of high redshift clusters in Section~\ref{sec_sample}. Theoretical predictions from $N$-body numerical simulations are discussed in Section~\ref{sec_cM}. The statistical analysis is detailed in Section~\ref{sec_analysis}. Main sources of systematic uncertainties are addressed in Section~\ref{sec_systematics}. The discussion of results is contained in Section~\ref{sec_discussion} whereas final considerations are in Section~\ref{sec_conclusions}.

Throughout the paper, we assume a flat $\Lambda$CDM cosmology with density parameters $\Omega_\mathrm{M}=0.3$, $\Omega_{\Lambda}=0.7$ and Hubble constant $H_0=100h~\mathrm{km~s}^{-1}\mathrm{Mpc}^{-1}$, $h=0.7$. At $z=1$, $1\arcsec$ corresponds to $5.6~\mathrm{kpc}/h$ ($\simeq8.0~\mathrm{kpc}$).

\section{Cluster sample}
\label{sec_sample}

\begin{table}
\centering
\begin{tabular}[c]{lcccl}
        \hline
        \noalign{\smallskip}
        name & $z$ & $z_\mathrm{arc}$ & $\theta_\mathrm{arc}$ [$\arcsec$]& ref.    \\
        \noalign{\smallskip}
        \hline
XMMXCS J2215-1738	&	1.46	&	--	&	--	&	1	\\
XMMU J2205-0159		&	1.12	&	--	&	--	&	1	\\
XMMU J1229+0151		&	0.98	&	--	&	--	&	1	\\
WARPS J1415+3612	&	1.03	& 3.898	&	6.7	&	1, 2	\\
ISCS J1432+3332		&	1.11	&	--	&	--	&	1	\\
ISCS J1429+3437		&	1.26	&	--	&	--	&	1	\\
ISCS J1434+3427		&	1.24	&	--	&	--	&	1	\\
ISCS J1432+3436		&	1.35	&	--	&	--	&	1	\\
ISCS J1434+3519		&	1.37	&	--	&	--	&	1	\\
ISCS J1438+3414		&	1.41	&	--	&	--	&	1	\\
RCS 0220-0333		&	1.03	&	--	&	--	&	1	\\
RCS 0221-0321		&	1.02	&	--	&	--	&	1	\\
RCS 0337-2844		&	1.1	&	--	&	--	&	1	\\
RCS 0439-2904		&	0.95	&	--	&	--	&	1	\\
RCS 2156-0448		&	1.07	&	--	&	4.3	&	1, 3	\\
RCS 1511+0903		&	0.97	&	--	&	--	&	1	\\
RCS 2345-3632		&	1.04	&	--	&	--	&	1	\\
RCS 2319+0038		&	0.91	& 3.860	&	12.9	&	1, 3	\\
XLSS J0223-0436		&	1.22	&	--	&	--	&	1	\\
RDCS J0849+4452		&	1.26	&	--	&	4.4	&	1, 4	\\
RDCS J0910+5422		&	1.11	&	--	&	--	&	1	\\
RDCS J1252-2927		&	1.24	&	--	&	16.0	&	1, 5	\\
XMM J2235-2557		&	1.39	&	3.3$^\mathrm{a}$	&	11.5	&	6	\\
CL J1226+3332		&	0.89	&	--	&	14.0	&	7	\\
MS 1054-0321			&	0.83	&	--	&	--	&	8	\\
CL J0152-1357			&	0.84	& 3.928	&	15.4	&	9, 10	\\
RDCS J0848+4453		&	1.27	&	--	&	--	&	4	\\
XMMU J1230.3+1339	&	0.97	&	--	&	--	&	11	\\
ClG J1604+4304		&	0.90	&	--	&	--	&	12	\\
RX J1716.6+6708		&	0.81	&	--	&	--	&	13	\\
ClG 1137.5+6625		&	0.78	&	--	&	14.4	&	13	\\
        \hline
       \end{tabular}
\caption{List of high-$z$ clusters of galaxies. Redshift and angular radius of strong lensing arcs are reported in col.~3 and 4, respectively. References are quoted in col.~5: (1) \citet{jee+al11}; (2) \citet{hua+al09}; (3) \citet{gla+al03}; (4) \citet{jee+al06}; (5) \citet{lid+al04}; (6) \citet{jee+al09}; (7) \citet{je+ty09}; (8) \citet{jee+al05b}; (9) \citet{jee+al05a}; (10) \citet{ume+al05}; (11) \citet{ler+al11}; (12) \citet{mar+al05}; (13) \citet{clo+al00}. $^\mathrm{a}$:photometric redshift.
}
\label{tab_sample}
\end{table}

We collected published data for 31 high redshift galaxy clusters at $0.8 \ls z\ls 1.5$ with weak-lensing analysis, see Table~\ref{tab_sample}. The clusters were discovered in different surveys and with different techniques, and do not represent a complete sample. Some of them were discovered in X-ray surveys, such as the ROSAT Deep Cluster Survey (RDCS), the XMM Cluster Survey (XCS), the XMM LSS survey, the XMM-Newton Distant Cluster Project (XDCP), and the Wide Angle ROSAT Pointed Survey (WARPS), other in optical surveys such as the Red-sequence Cluster Survey-I (RDCS) and the Spitzer IRAC Shallow Survey (ISCS). Details on the clusters and the surveys can be found in the papers quoted in Table~\ref{tab_sample}. 

In absence of new data, we based our study on already published tangential shear profiles. Radial profiles usually extend to either $\ls 2\arcmin$ for space based observations or $\gs 3\arcmin$ for ground-based observations with the notable exception of XMMU J1230.3+1339, whose profile extends to $\sim 6\arcmin$. We refer to the original papers for details on the background galaxy selection, the shear measurement and the estimation of the source depth. 

A sub-sample of eight galaxy clusters shows an evident gravitational arc\footnote{The galaxy cluster RCS~J0220.9-0333 also shows a gravitational arc in the HST archival images, but we do not classify this as a cluster-scale lensing system, because of the relatively small radius ($\sim 1.4\arcsec$) of the arc and its location (centred around the second brightest galaxy in the cluster).},
whose radius and redshift (when available) are reported in Table~\ref{tab_sample}. We used the strong lensing information to determine the cluster total mass at the Einstein radius, see Section~\ref{sec_analysis}. 

We also collected available information from literature in order to define the dynamical state of the systems. Because of paucity of observational information (i.e., deep X-ray and/or optical imaging, complete spectroscopic surveys and so on), we can robustly asses the dynamical state only for nine clusters, according to the presence of sub-clumps or filaments (as seen in the HST or X-ray images), the velocity distribution of galaxies and the position of the brightest cluster galaxy relatively to the X-ray emission peak. 

Seven galaxy clusters show clear deviations from equilibrium (XMMXCS J2215-1738, XMMU J2205-0159, RCS 0439-2904, CL J1226+3332, MS 1054-0321, CL J0152-1357, RX J1716.6+6708) and two appear to be close to the hydrodynamical equilibrium (ISCS J1432+3332, ISCS J1438+3414). The predominance of out-of-equilibrium systems is clearly expected, as massive galaxy clusters at high-redshift are likely observed in an active assembling phase. For the majority of the systems, we decided not to over-interpret the available information (usually, relatively shallow mass density maps derived with weak lensing analyses and/or X-ray/optical imaging) and not to classify them.

\section{Theoretical predictions}
\label{sec_cM}

Dark matter haloes are well described as NFW density profiles with  \citep{nfw96,nav+al97,ji+su02},
\begin{equation}
\label{nfw1}
	\rho_\mathrm{NFW}=\frac{\rho_\mathrm{s}}{(r/r_\mathrm{s})(1+r/r_\mathrm{s})^2}.
\end{equation}
The radius $r_{200}$ is defined such that the mean density contained within a sphere of radius $r_{200}$ is 200 times the critical density at the halo redshift. The corresponding concentration is $c_{200} \equiv r_{200}/ r_\mathrm{s}$. $M_{200}$ is the mass of such sphere.

$N$-body simulations \citep{og+bl09,mac+al08,gao+al08,duf+al08,pra+al11} have been providing a useful picture of the expected properties of dark matter haloes. Results may depend on the overall normalisation of the power spectrum, the mass resolution, the simulation volume \citep{pra+al11} or the binning and fitting procedures \citep{me+ra13}. 

The dependence of halo concentration on mass and redshift is usually described by a power law,
\beq
\label{eq_cM_1}
c =A(M/M_\mathrm{pivot})^B(1+z)^C.
\eeq
At low masses and redshifts, the power-law modelling provides an excellent fit to dark matter simulated haloes \citep{net+al07,mac+al08,gao+al08,duf+al08}. As reference, we follow \citet{duf+al08}, who used the cosmological parameters from WMAP5 and found $\{A,B,C\}=\{ 5.71 \pm0.12, -0.084 \pm 0.006, -0.47\pm0.04\}$ for a pivotal mass $M_\mathrm{pivot}=2\times10^{12}M_\odot/h$ in the redshift range $0-2$ for their full sample of clusters. 

Recently, \citet{pra+al11} claimed that the concentration and its evolution depend on the root-mean-square fluctuation amplitude of the linear density field. They noticed a flattening and upturn of the relation with increasing mass and estimated concentrations for galaxy clusters substantially larger than results of \citet{duf+al08}.

This $c(M)$ relation, presented in \citet[ equations~12--23]{pra+al11}, provides a functional form showing evolution with redshift and mass alternative to the simple featureless power-law in Eq.~(\ref{eq_cM_1}). Even if it is still questioned whether this relation is fit to describe all the clusters at a given redshift \citep{me+ra13}, it might catch important features regarding at least unrelaxed clusters \citep{lud+al12}. Since large values of $c_{200}$ may emerge from both observation and simulations, we will also consider a $c(M)$ relation favouring higher concentrations and modelled as suggested in \citet{pra+al11} as an alternative relation.

The scatter in the concentration about the median $c(M)$ relation is lognormal \citep{duf+al08,bha+al11},
\beq
\label{eq_cM_2}
p(\ln c | M)=\frac{1}{\sigma\sqrt{2\pi}}\exp \left[ -\frac{1}{2} \left(  \frac{\ln c - \ln c(M)}{\sigma}\right) \right],
\eeq
with a dispersion $\sigma (\log c_{200})=0.15$ for a full sample of clusters \citep{duf+al08}.

\section{Data analysis}
\label{sec_analysis}

\begin{table*}
\centering
\begin{tabular}[c]{lr@{$\,\pm\,$}lr@{$\,\pm\,$}lccc}
        \hline
        \noalign{\smallskip}
	name	& \multicolumn{2}{c}{$M_{200}$} & \multicolumn{2}{c}{$c_{200}$} & $P_\mathrm{straight}$& $P_\mathrm{over-concentrated}$&$P_\mathrm{unrelated}$    \\
        \noalign{\smallskip}
        \hline
XMMXCS J2215-1738	&	2.0	&	1.6	&	4.5	&	5.1	&	0.38	&	0.35	&	0.28	\\
XMMU J2205-0159	&	1.4	&	1.0	&	3.4	&	4.3	&	0.38	&	0.34	&	0.27	\\
XMMU J1229+0151	&	7.6	&	5.1	&	1.4	&	1.0	&	0.16	&	0.55	&	0.29	\\
WARPS J1415+3612	&	2.2	&	1.4	&	3.9	&	2.0	&	0.43	&	0.43	&	0.15	\\
ISCS J1432+3332	&	5.5	&	4.5	&	1.8	&	1.4	&	0.20	&	0.57	&	0.23	\\
ISCS J1429+3437	&	5.8	&	4.6	&	1.3	&	1.4	&	0.20	&	0.45	&	0.35	\\
ISCS J1434+3427	&	1.8	&	1.9	&	1.5	&	2.3	&	0.31	&	0.35	&	0.34	\\
ISCS J1432+3436	&	3.0	&	2.8	&	2.2	&	2.8	&	0.29	&	0.40	&	0.31	\\
ISCS J1434+3519	&	2.6	&	3.0	&	0.9	&	1.2	&	0.25	&	0.35	&	0.40	\\
ISCS J1438+3414	&	1.6	&	1.6	&	2.2	&	3.2	&	0.34	&	0.36	&	0.30	\\
RCS 0220-0333	&	3.0	&	1.8	&	4.2	&	3.5	&	0.43	&	0.38	&	0.19	\\
RCS 0221-0321	&	1.0	&	0.5	&	9.2	&	5.7	&	0.46	&	0.25	&	0.29	\\
RCS 0337-2844	&	4.8	&	4.9	&	2.1	&	2.5	&	0.32	&	0.43	&	0.25	\\
RCS 0439-2904	&	2.4	&	1.2	&	5.1	&	3.8	&	0.48	&	0.33	&	0.20	\\
RCS 2156-0448	&	0.6	&	0.7	&	5.1	&	5.5	&	0.41	&	0.34	&	0.26	\\
RCS 1511+0903	&	1.0	&	0.6	&	7.2	&	6.0	&	0.42	&	0.30	&	0.28	\\
RCS 2345-3632	&	1.8	&	1.6	&	1.6	&	1.8	&	0.30	&	0.45	&	0.26	\\
RCS 2319+0038	&	1.7	&	0.7	&	10.5	&	4.8	&	0.54	&	0.16	&	0.30	\\
XLSS J0223-0436	&	8.1	&	5.2	&	1.5	&	1.3	&	0.19	&	0.52	&	0.29	\\
RDCS J0849+4452	&	2.6	&	1.3	&	2.4	&	1.1	&	0.20	&	0.63	&	0.17	\\
RDCS J0910+5422	&	2.5	&	0.9	&	5.6	&	3.0	&	0.55	&	0.27	&	0.18	\\
RDCS J1252-2927	&	4.6	&	1.2	&	4.0	&	1.6	&	0.41	&	0.44	&	0.15	\\
XMM J2235-2557	&	3.8	&	1.5	&	5.2	&	2.3	&	0.56	&	0.26	&	0.18	\\
CL J1226+3332	&	7.0	&	1.7	&	3.6	&	0.9	&	0.33	&	0.53	&	0.14	\\
MS 1054-0321	&	5.1	&	1.5	&	7.0	&	5.0	&	0.49	&	0.31	&	0.21	\\
CL J0152-1357	&	2.2	&	0.4	&	10.6	&	3.6	&	0.57	&	0.05	&	0.38	\\
RDCS J0848+4453	&	1.7	&	2.0	&	0.9	&	1.3	&	0.26	&	0.33	&	0.41	\\
XMMU J1230.3+1339	&	8.5	&	5.7	&	2.0	&	2.6	&	0.26	&	0.45	&	0.29	\\
ClG J1604+4304	&	2.2	&	2.1	&	3.9	&	5.3	&	0.38	&	0.33	&	0.29	\\
RX J1716.6+6708	&	3.7	&	2.4	&	3.6	&	3.5	&	0.38	&	0.40	&	0.22	\\
ClG 1137.5+6625	&	7.2	&	3.2	&	3.1	&	1.4	&	0.31	&	0.55	&	0.14	\\	
\hline
	\end{tabular}
\caption{Masses (in units of $10^{14}M_\odot/h$) and concentrations of the high-$z$ clusters determined assuming priors uniform in logarithmically spaced intervals.  $P_\mathrm{straight}$ ($P_\mathrm{over-concentrated}$)  is the probability that the $c(M)$ relation determined in \citet{duf+al08} \citep{pra+al11} is true. $P_\mathrm{unrelated}$ is the probability that mass and concentration are not related.
}
\label{tab_M200_c200}
\end{table*}

\begin{figure}
       \resizebox{\hsize}{!}{\includegraphics{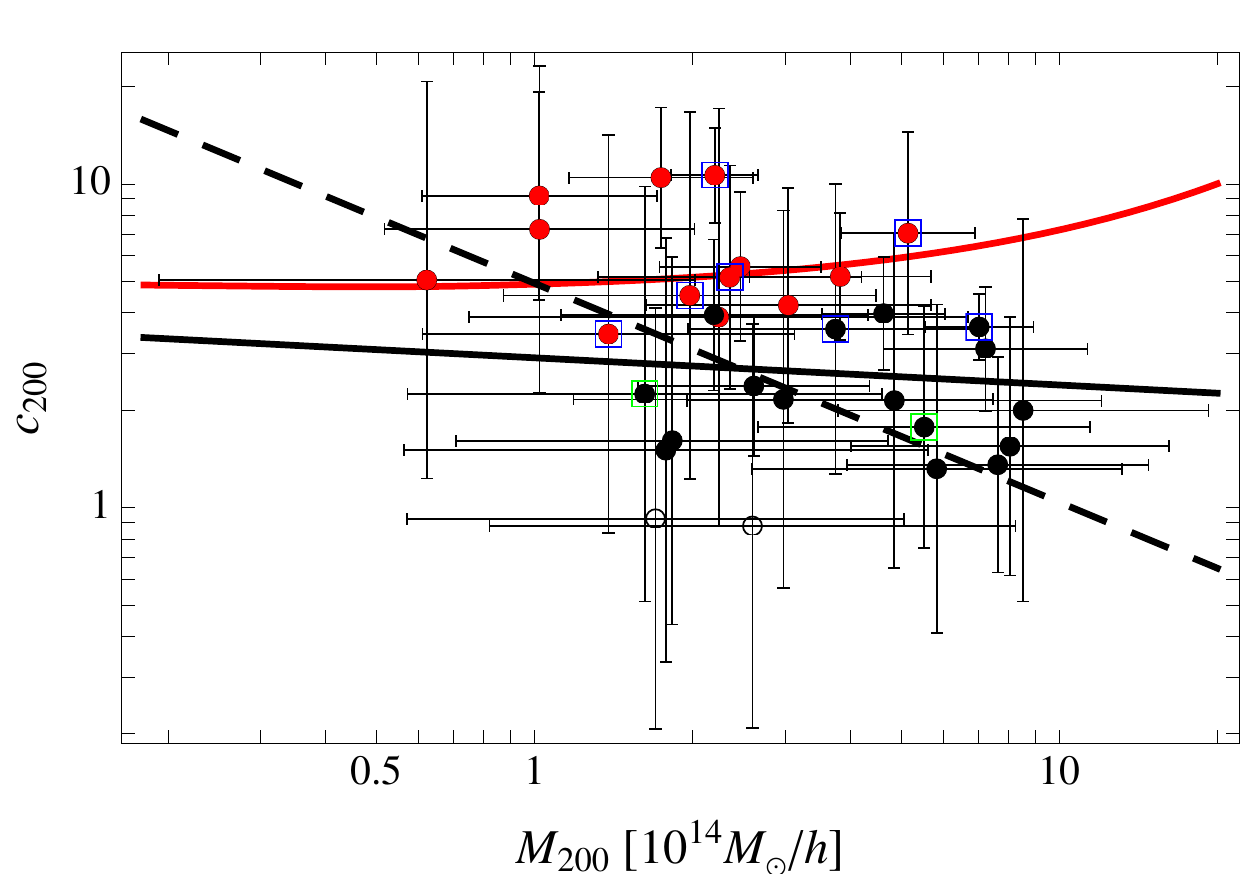}}
       \caption{Masses and concentrations for the high-$z$ clusters. $M_{200}$ and $c_{200}$ were determined assuming priors uniform in logarithmically spaced intervals. The black line is the $c(M)$ power law determined in \citet{duf+al08} for $z\simeq 1.1$, i.e., the mean redshift of the cluster sample; the red line is the $c(M)$ relation found in \citet{pra+al11}. The dashed black line is the best-fitting relation for $C=0$. Concentrations of clusters marked as black (red) disks are better explained by the $c(M)$ of \citet{duf+al08} (\citet{pra+al11}, respectively). Empty circles marks clusters where the unrelated hypothesis is preferred. Blue (green) squares denote irregular (regular) clusters.}
	\label{fig_M200_c200}
\end{figure}

\begin{figure}
       \resizebox{\hsize}{!}{\includegraphics{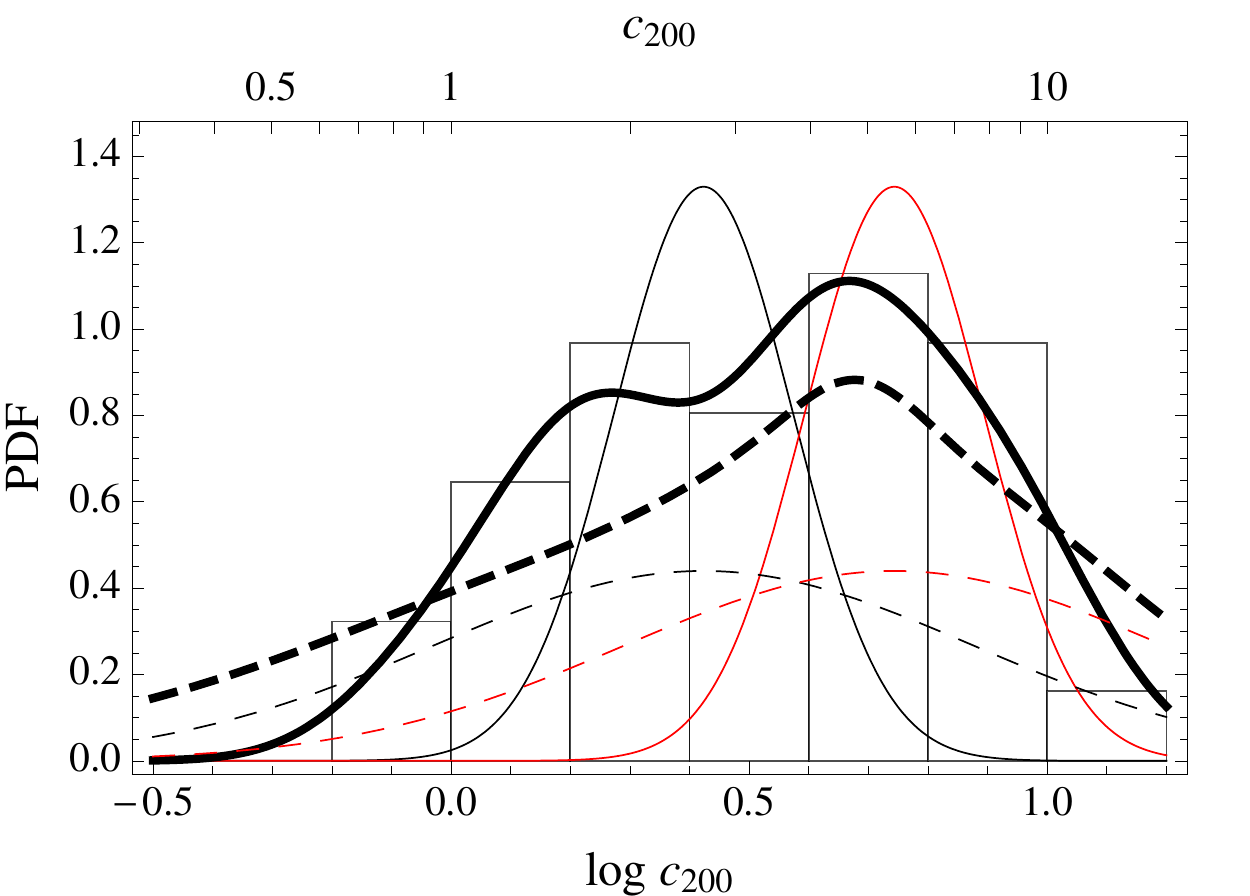}}
       \caption{The one-dimensional normalised distribution of measured concentrations. The thick line is the observed distribution of the central values smoothed using a Gaussian kernel estimator; the red (thin black) line is the lognormal distribution in the over-concentrated (straight) case, normalised to 1/2, for the average mass $\simeq 3.6\times 10^{14} M_\odot/h$ at $z \simeq 1.1$. Dashed lines represent the corresponding distributions broadened by the observational uncertainties.}
	\label{fig_histo_c200}
\end{figure}

To assess realistic probability distributions for the halo parameters we performed a statistical Bayesian analysis, as already done in a number of gravitational lensing studies \citep{ogu+al05,cor+al09,se+um11}. The Bayes theorem states that \citep{mac03}
\beq
\label{baye}
p(\bfP | \bfd) \propto {\cal L}( \bfP|\bfd) p(\bfP),
\eeq
where $p(\bfP | \bfd)$ is the posterior probability of the parameters $\bfP$ given the data $\bfd$, ${\cal L}( \bfP|\bfd)$ is the likelihood of the model parameters given the data, and $p(\bfP)$ is the prior probability distribution for the model parameters.

We constrained mass and concentration with reduced shear profiles from weak-lensing analyses and, when available, the strong lensing constraint on the Einstein radius. The combined $\chi^2$ function is defined as \citep{um+br08}
\beq
\label{eq_chi_2}
\chi^2=\sum_i \frac{\left[g_{+}(\theta_i)-g_{+}^\mathrm{NFW}(\theta_i)\right]^2}{\sigma_{+}^2(\theta_i)}+
\frac{\left[ 1-g_{+}^\mathrm{NFW}(\theta_\mathrm{E},z_\mathrm{E})\right]^2}{\sigma_{+}^2(\theta_\mathrm{E})}.
\eeq
The first term accounts for the reduced tangential shear $g_{+}$ measured in circular annuli at angular position $\theta_i$. The second term is the strong lensing contribution and exploits the fact that the reduced shear equals one at the Einstein radius. In Eq.~(\ref{eq_chi_2}), $g_{+}$ is the measured value and $g_{+}^\mathrm{NFW}(\theta; M_{200},c_{200})$ is the theoretical prediction for a NFW lens at radius $\theta$ \citep{wr+br00}. The related likelihood is ${\cal L} \propto \exp{-\chi^2/2}$. 

Basic strong lensing analyses can provide accurate measurements of the total projected mass within giant arcs even for very irregular clusters, whereas they may fail in constraining the profile. The determination of the Einstein radius is then reliable notwithstanding the possible complex structure of the cluster core and can be used as a constraint for the shear profile \citep{um+br08,ogu+al12}. We assume conservatively a 10 per cent error for the Einstein radius if the arc redshift has been determined spectroscopically. This error accounts for uncertainties related to the lens centre determination and the spherical approximation and is much larger than the actual image angular resolution. 

If the redshift of the arc is unknown, we fix $z_\mathrm{arc}$ at the mean of the measured arc redshifts in the sample and adopt a larger error of 20 per cent. We then propagate this uncertainty to $g_{+}$ assuming a singular isothermal sphere, so that $\sigma_{+}(\theta_\mathrm{E})=2\sigma_{\theta_E}/\theta_E$ \citep{um+br08}. 

Through the priors, we can test competing proposals for the $c(M)$ relation. For the mass, we always used a prior uniform in logarithmically spaced intervals between $10^{13}$ and $2\times 10^{15}M_\odot h^{-1}$ and null otherwise. For the concentration, we studied three cases. Firstly, we did not consider any relation between mass and concentration and let $c_{200}$ be uniformly distributed in logarithmically spaced intervals between 0.1 and 20 and null otherwise (the ``unrelated" case). This prior is not informative, so that any trend on concentration recovered under this assumption is attributable to the data. Secondly, we used a log-normal distribution, Eq.~(\ref{eq_cM_2}), with the $c(M)$ relation determined in \citet{duf+al08} (the ``straight" case). Thirdly, a log-normal distribution with the $c(M)$ relation determined in \citet{pra+al11} (the ``over-concentrated" case). 

The effect of alternative uniform priors on mass and concentration is discussed in Section~\ref{sec_syst_priors}.

Masses and concentrations are listed in Table~\ref{tab_M200_c200} and plotted in Figure~\ref{fig_M200_c200}. These values were determined under the less informative priors, i.e., in the ``unrelated" case. The one-dimensional distribution of concentrations is shown in Figure~\ref{fig_histo_c200}. The observed distribution was compared with the lognormal distributions in the over-concentrated case (red line) and the straight case (thin black line), both normalised to 1/2, for a cluster mass equal to the average of the sample, $\langle M_{200}\rangle \simeq 3.6\times 10^{14} M_\odot/h$, at the average redshift, $z\simeq1.1$. There is a peak in concentrations around the maximum of the over-concentrated case but, given the very large uncertainties, distributions are very broad and we cannot draw sensible conclusions from the one-dimensional analysis alone.

\subsection{Evolution}

Let us consider the evolution of the concentration with mass and redshift. Figure~\ref{fig_M200_c200} suggests that measured concentrations are correlated with the mass. More massive clusters have on average smaller concentrations. To quantify this effect, we fitted the $c(M,z)$ relation with a power law as in Equation~(\ref{eq_cM_1}). The parameters to determine are the normalisation, the mass slope and the redshift slope. We fixed the pivotal mass to $2.5 \times 10^{14}M_\odot h^{-1}$, nearby the median mass of the cluster sample. As $\chi^2$ function, we used
\beq
\chi^2=\sum_i \frac{\left[ (\log c_{200})^\mathrm{obs}_i-\log c_{200}(M_{200,i})\right]^2}{(\delta_{\log c_{200}})_i^2+\left| \frac{\partial \log c_{200}}{\partial \log M_{200}}\right|^2(\delta_{\log M_{200}})_i^2 +\sigma^2_{\log c_{200}}},
\eeq
where the error on the mass is dealt with according to \citet{dag05}. The intrinsic scatter $\sigma_{\log c_{200}}$ is fixed to 0.15 \citep{duf+al08}. We neglected errors on cluster redshift determinations. Parameter space was explored by running Montecarlo Markov chains to recover the full posterior probabilities. 

We found $\log A = 1.69 (1.67)\pm 0.61$, $B=-0.83 (-0.71) \pm0.39$ and $C=-3.2 (-3.2)\pm1.9$. Central values and dispersions are the bi-weight estimators of the marginalised distributions. Best-fitting values are reported in parentheses. Results are stable against the exclusion of arcs, i.e., neglecting the strong lensing constraints in Eq.~(\ref{eq_chi_2}). With a weak lensing only analysis, we got $\log A = 1.78\pm 0.66$, $B=-0.65 \pm0.40$ and $C=-3.4\pm2.1$.

The measured mass slope is much steeper than theoretical expectations. The parameter $B$ is smaller than $-0.084$, the value predicted by \citet{duf+al08}, at the 99.1 per cent confidence level. We find good agreement with other observational results at lower redshifts. \citet{ogu+al12} found $B=-0.59 \pm0.12$ in a strong-lensing selected sample of massive clusters in the redshift range $0.3\ls z \ls 0.7$. \citet{oka+al10} analysed 19 X-ray luminous lensing galaxy clusters at $0.15\ls z\ls 0.3$ and found $B=-0.40 \pm0.19$.

We also found some suggestions for clusters at higher $z$ being less concentrated. $C$ is not positive at the 95.8 per cent confidence level. We caution that this trend might be an artefact in the sample selection whether over-concentrated clusters at lower redshift were preferentially included. Since the very different selection methods used to build our sample, this kind of bias is difficult to asses.

The main dependence of observed concentrations is on mass. If we ignore the redshift dependence of the mass-concentration relation as expected over the redshift range covered by our sample \citep{net+al07,duf+al08} and fix $C=0$, we find $B=-0.78(-0.68)\pm0.40$. This result is not distinguishable from the case with allowed redshift evolution.

\subsection{Stacked analysis}
\label{sec_stack}

\begin{figure}
       \resizebox{\hsize}{!}{\includegraphics{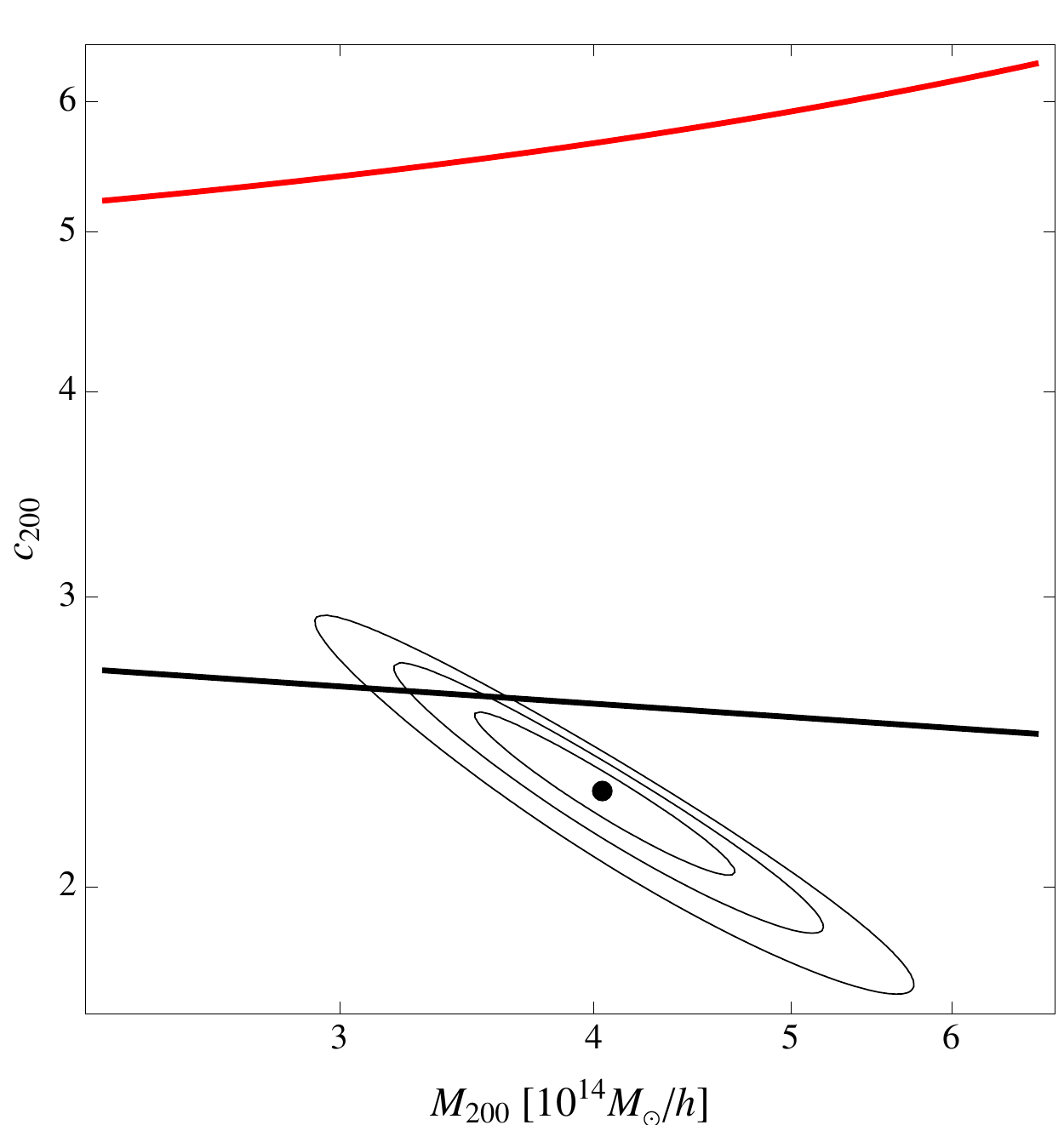}}
       \caption{Posterior probability distribution for mass and concentration derived with the stacked analysis. The dot marks the model with the max probability. Contours are plotted at fraction values $\exp (-2.3/2)$, $\exp(-6.17/2)$, and $\exp(-11.8/2)$ of the maximum, which denote confidence limit regions of 1, 2 and $3\sigma$ in a maximum likelihood investigation, respectively. The black line is the $c(M)$ power law determined in \citet{duf+al08} for $z\simeq 1.1$, i.e., the mean redshift of the cluster sample; the red line is the $c(M)$ relation found in \citet{pra+al11}.}
	\label{fig_M200_c200_stacked}
\end{figure}

A usual technique to enhance the weak lensing signal is to stack shear measurements from numerous clusters \citep{joh+al07,man+al08,ogu+al12,oka+al13}. To study the mean properties of the sample, we adopted an alternative approach. We fitted all clusters at once assuming that all of them have the same mass and concentration. The related $\chi^2$ function is then
\beq
\chi^2_\mathrm{Stacked}=\sum_j \chi^2_j (z_j; M_{200}, c_{200})
\eeq
where the sum extends over the sample and $\chi^2_j$ is the expression defined in Eq.~(\ref{eq_chi_2}) for the $j$th cluster. To compute $\chi^2_j$ we have to consider a NFW halo at the redshift of the $j$th cluster.

This approach has some advantage with respect to usual stacking methods: $i$) the redshift dependence is completely removed; $ii$) the use of the reduced shear instead of the shear allows us to extend the analysis even in the inner regions; $iii$) we do not have to choose to stack either in physical or rescaled radius, which can artificially make the stacked profile shallower.

As for individual clusters, $M_{200}$ and $c_{200}$ were determined assuming priors uniform in logarithmically spaced intervals. For mass and concentration we found $M_{200}=(4.1\pm0.4)\times10^{14}M_\odot/h$ and $c_{200}=2.26\pm0.17$ in good agreement with the prediction from \citet{duf+al08} at the average redshift $\sim 1.1$. The posterior probability function is plotted in Fig.~\ref{fig_M200_c200_stacked}. The reduced $\chi^2$ is $\simeq 2.1$ ($\chi^2\simeq436$ for 205 data points and two free parameters). This does not account for the intrinsic scatter in the $c(M)$ relation, $\sigma_{\log c_{200}}\simeq 0.15$. For a NFW halo of $\sim 4\times10^{14}M_\odot/h$ at $z=1.1$, the scatter in concentration can be viewed as an intrinsic error on the shear profile of $\sigma_+\simeq0.01$ ($0.6\times10^{-2}$) at $\theta \simeq 50\arcsec$ ($100\arcsec$), which brings the reduced $\chi^2$ to $\ls 1.1$.

As already found in similar analyses at lower redshift \citep{joh+al07,man+al08,ogu+al12,oka+al13}, there is no over-concentration problem whether we assume that all clusters follows the same profile.

\subsection{Sub-groups in the sample}

The analysis of the whole sample put in evidence some correlations between concentration, mass and redshift. We can further investigate the properties of the lenses by considering whether the clusters can be divided in homogeneous sub-samples. To this aim, we considered the three scenarios for the $c(M,z)$ relation discussed above, i.e., the unrelated, the straight and the over-concentrated case, and how well the mass and the concentration of each cluster can be explained by each case. 

We see by eye that nearly one half of the sample in Figure~\ref{fig_M200_c200}, mostly the lower mass clusters, better follows the over-concentrated relation. The other half is better interpreted in terms of a power-law $c(M)$ with low normalisation. 

In order to quantify and back up this impression, we estimated the probabilities for different scenarios by computing Bayesian evidences \citep{mac03}. To compute the Bayesian evidence, we had to integrate the posterior probability distribution $p(M_{200},c_{200}|\bfd)$, so that the degeneracy between the two parameters, the statistical uncertainties and all minor details are considered in the estimate of the probability of each scenario. The scatter in the theoretical $c(M)$ relations is included as well. The straight, over-concentrated and unrelated cases were considered as equiprobable a priori. Probabilities are listed in Table~\ref{tab_M200_c200}. In this context, the non-monotonic feature in the over-concentrated case is irrelevant due to the large errors on mass and concentration. Any $c(M)$ relation with a high normalisation would have implied similar results. 

Differences in probability among competing scenarios for each cluster are not very large due to the large statistical errors of the shear profiles and the related broad probability distribution $p(M_{200},c_{200}|\bfd)$. On the other side, the low accuracy in the evaluation of each cluster is compensated by the quite large number of high-$z$ clusters we considered.
 
Bayesian evidences for single clusters can be combined to compute the total probability that a sample is drawn from a given population as opposed to a collection with different intrinsic properties. If we assume that all clusters are representative of a single population featuring a given $c(M)$ relation, the probability of the straight case is of $\sim 78\%$ and exceeds the $\sim 22\%$ probability for the over-concentrated case. This confirms the results of the stacked analysis in Section~\ref{sec_stack} in favour of the theoretical predictions by \citet{duf+al08}.

On a cluster by cluster basis, the scenario with an over-concentrated $c(M)$ is preferred by 13 out of 31 clusters. For a narrow majority of 16 clusters, the probability is maximised in the straight case. Finally, the unrelated case has the larger statistical probability for the 2 clusters with the lower concentration, $c_{200}\sim 1$. 

An alternative point of view is that clusters are not representative of a single $c(M)$ relation but may instead be separated according to their properties.
Let us compare the simple case in which all clusters follow the straight relation to a more complicated scenario in which haloes are a mixture of either regular or intrinsically over-concentrated clusters. This distribution can be approximated by assuming that one half of the clusters of a given mass follows the straight relation and that the second half follows the over-concentrated relation. Occam's razor suggests that simpler models should have higher prior probability. One sensible recipe is that the prior probability for a scenario with $N_\mathrm{pop}$ distinct populations scales as $(1/2)^{N_\mathrm{pop}+1}$ \citep{got+al01}. A priori the scenario with only one population is then two times more likely than the mixture of two kinds. The posterior probability for the single population case is of $\sim 42\%$, marginally lower than the $\sim 58\%$ probability for the mixture.

Mass and concentration of clusters showing sign of dynamical activity seem to be over-concentrated, see Figure~\ref{fig_M200_c200}. We considered two distinct hypotheses, i.e., that clusters are members of a population following either the over-concentrated or the straight relation. The probability of the seven irregular clusters, see Section~\ref{sec_sample}, to be from a population of haloes following the over-concentrated (straight) relation is of $\sim$94.6 (5.4) per cent. On the opposite side, the two clusters with signs of regularity prefer the straight (over-concentrated) relation at the $\sim$74.8 (25.2) per cent. 

The evolution in mass and redshift can be detected only considering the full sample which contains both the higher concentration, lower mass clusters which favour the over-concentrated relation and the lower concentration, higher mass haloes which are in line with the straight case. Attempting to fit a power-law $c(M,z)$ to the two subsamples separately is unsuccessful in constraining the slope parameters $B$ and $C$ due to the large statistical errors and the smaller number of points.

\section{Systematics}
\label{sec_systematics}

We discuss the main sources of systematic uncertainties that might plague our analysis.

\subsection{Triaxiality}

\begin{table}
\centering
\begin{tabular}[c]{lr@{$\,\pm\,$}lc}
        \hline
        \noalign{\smallskip}
        name & \multicolumn{2}{c}{$e_\Delta^\mathrm{gas}$} & ref.   \\
        \noalign{\smallskip}
        \hline
XMMXCS J2215-1738	&	 \multicolumn{2}{c}{$\gs1.30$} 	&	1	\\
WARPS J1415+3612	&	0.62	&	0.33	&	1	\\
RDCS J0849+4452	&	 \multicolumn{2}{c}{$\gs0.21$}	&	1	\\
RDCS J0910+5422	&	 \multicolumn{2}{c}{$\gs1.16$}	&	1	\\
RDCS J1252-2927	&	 \multicolumn{2}{c}{$\gs0.41$}	&	1	\\
XMM J2235-2557	&	1.49	&	0.57	&	1	\\
CL J1226+3332	&	1.17	&	0.35	&	2	\\
MS 1054-0321		&	0.97	&	0.29	&	2	\\
RX J1716.6+6708	&	0.70	&	0.36	&	2	\\
ClG 1137.5+6625	&	0.55	&	0.19	&	2	\\
        \hline
       \end{tabular}
\caption{Elongations of the gas distribution for a subsample of galaxy clusters. References: (1) \citet{cul+al10}; (2) \citet{ree+al10} 
}
\label{tab_elongation}
\end{table}

One of the main causes of systematic errors in the determination of mass and concentration from weak lensing is triaxiality \citep{ras+al12}. Lensing clusters are preferentially selected if elongated along the line of sight. Mass and concentration may be consequently under or over-estimated. Complete multi-wavelength data-sets from X-ray to optical to SZ effect  which are needed to perform unbiased triaxial analyses \citep{lim+al12} are still not available for the clusters studied in this paper. We anyway searched for data in literature in order to constrain the gas distribution. Even if the gas may not follow the total matter distribution probed by lensing, especially in merging and irregular clusters, we still expect some correlation. 

We applied the method first used in \citet{def+al05,ser+al06} and then refined in \citet{ser+al12a} to a sub-sample of 10 clusters with X-ray and SZ data. The effect of orientation and triaxiality can be parameterised by the parameter $e_\Delta$, i.e., the ratio between the width of the cluster in the plane of the sky to the size along the line of sight \citep{ser07}. If $e_\Delta <(>)1$, the cluster is elongated toward the observer (in the plane of the sky). Results are listed in Table~\ref{tab_elongation}. We found no evidence for orientation bias, with nearly one half of the clusters preferentially elongated along the line of sight and the other half in the plane of the sky.

Even if we have no clear clue for any effective shape or orientation bias in our sample, we can anyway test how much this bias could affect our results. Weak lensing data for our sample are not deep enough to detect ellipticity and orientation in the projected mass distribution, and the circular approximation is used to fit the data \citep{jee+al11}. The bias can then be estimated by assuming that each cluster is either prolate or oblate, with the symmetry axis oriented along the line of sight. The cluster is therefore still circular in projection.

According to the analysis in \citet{ji+su02}, a cluster of $\sim 3\times 10^{14}M_\odot/h$ at $z\sim1$ should have axial ratios of $\sim 0.4$ (minor-to major axis length) and $\sim 0.5$ (intermediate-to major axis length). If oriented along the line of sight, as expected for lensing selected clusters, this would imply an elongation parameter of $e_\Delta \sim 0.5$. We then re-did our analysis under the hypothesis that all clusters in the sample are prolate and oriented along the line of sight with an intrinsic axial ratio of $e_\Delta=0.5$. We exploited the formalism for lensing by ellipsoidal NFW haloes detailed in \citet{ser+al10,ser+al10b,se+um11}. The main trends detected in the $c(M)$ relation are not affected. On average masses and concentrations are lower by $25\pm0.9$ and $26\pm17$ per cent, respectively. The over-concentrated $c(M)$ is preferred by 11 clusters. The straight case has the maximum probability in 13 cases. The unrelated case is preferred in the remaining 7 clusters.

\subsection{Substructures and large-scale structure}

Substructures close to the line of sight may inflate concentrations. The unknown location of the substructure along the line of sight also makes the weak lensing mass estimate problematic when fitting the reduced tangential shear profile with a NFW profile. Due to substructures, concentrations are on average biased low by  $\sim$10 per cent \citep{men+al10,gio+al12}. This bias increases with mass, since most massive haloes have a larger fraction of their mass in substructures \citep{gio+al12}. In case of massive clumps just outside the virial radius of the cluster, the tangential shear is diluted and the mass profile is severely under-estimated \citep{men+al10}. \citet{be+kr11} found that correlated large-scale structure within several virial radii of clusters may contribute a significant amount to the scatter in the mass determination, up to 20 per cent for massive clusters at $z \ls 0.5$.

Even if a number of clusters in our sample shows substructures and signs of merging activity or belong to large superclusters, due to the sparsity of spectroscopic data there is indication of sub-clumps along the line of sight for just a couple of them. RCS~0439-2904 might be a line of sight superposition of multiple clusters as suggested by the several components in the velocity histogram \citep{cai+al08}. A possible group at $z = 0.74$ composed of 33 members might be in front of RDCS~J1252-2927, leading to the overestimation of the WL mass \citep{lom+al05}. 

Any complex matter structure might challenge the simple spherical modelling. However, at large radii the mass profile is smooth for all clusters in the sample and given the large statistical uncertainties the reduced $\chi^2$ for the NFW fitting procedure is $\ls 1$ even for dynamical active clusters such as MS 1054-0321. The effect of cluster substructures and off-centring is generally minimised by removing from the analysis the inner radial bins in the shear profiles.

Since the discussed effects act in different directions and are smaller than the statistical uncertainties, we can then conclude that substructures are more likely to increase the scatter in our fits than to bias the results.

\subsection{Signal dilution and redshift uncertainty}

Dilution of shear signal and source redshift uncertainty are two of the main sources of systematics error in the interpretation of the measured shear profile. Shear signal may be diluted by inefficient selections of background galaxies. Without adequate colour information, the weak-lensing signal can be weakened particularly toward the cluster centre by the inclusion of cluster members in the background galaxy sample \citep{bro+al05b}, leading to an underestimation of the concentration \citep{lim+al07,um+br08} and of the central cluster mass \citep{oka+al10}.

Colour cuts and/or photometric redshift classifications to separate background galaxies from cluster members and foreground galaxies reduce the level of contamination to $\ls 10$ per cent at $r_{2500}$ (which roughly corresponds to $\sim1\arcmin$ for massive clusters at $z\sim1$), and to few per cents at larger radii \citep{hoe07}. The related error on mass is $\sim 5$ per cent \citep{hoe07}. Dilution effects can be further reduced by removing the innermost radial bin in the reduced shear profile, which we did whenever suggested in the original papers quoted in Table~\ref{tab_sample}.

The total error in the source redshift estimation due to sample variance, resampling error, and differences among the photometric redshift estimation codes is of order of 0.06, which in turn causes a $\sim 3\%$ ($\sim11\%$) uncertainty in mass for a cluster at $z\sim0.9$ ($1.4$) cluster \citep{jee+al09,jee+al11}.

These systematic errors are then negligible with respect to the statistical uncertainties in our sample, which are usually in excess of 50 per cent, see Section~\ref{sec_analysis}.

\subsection{Sample selection}

Due to paucity of high-$z$ clusters with lensing analyses, we had to consider all clusters we found in literature and we could not implement a well-defined selection function. The clusters in our sample were mostly discovered within X-ray or optical surveys. The sample is then not statistical, which might bias the results. However, some reassuring considerations can still be made.

The sample is numerous enough. Since the finding techniques were various, we do not expect that all discoveries were affected by the same bias, whereas we expect that biases due to the orientation and internal structure of clusters and the projection effect of large-scale structure are strongly mitigated for a large sample. On the other hand, sample inhomogeneity might increase the observed scatter.

The clusters were not originally selected with lensing methods. The lensing signal for each cluster was detected after the independent discovery. The sample is then not expected to be strongly affected by biases plaguing lensing selected samples, such as the over-concentration problem and the orientation bias \citep{og+bl09}. In particular, the orientation bias is limited for X-ray selected clusters \citep{men+al11,se+zi12}. 

Richness detection techniques exploits quite large angular areas in the sky. Counting of galaxies is affected by projection effects mainly in the inner cluster regions. If the area is large enough, the richness is less affected by orientation biases.

Being detected at such high redshift, the clusters are expected to be massive. However, this is appropriate for our study of the high mass end of the $c(M)$ relation. Since high concentrations enhance the signal in the inner regions, over-concentrated clusters might be preferentially included by detection methods focusing on the inner regions. This is not the case for our sample.

\subsection{Priors}
\label{sec_syst_priors}

The main risk of using priors is that if the likelihood is not peaked, final results might reflect the employed a priori hypotheses more than the data. To caution against this, we used not informative priors. Distributions uniform in logarithmically spaced decades are standard choices as priors for positive parameters such as mass and concentration. Flat priors in the allowed range provide an alternative. To check the effect of priors we re-did our analysis with this second choice. 

The main effect of uniform priors is to favour slightly larger concentrations. There is no bias in the mass estimate (the variation is of $0\pm14$ per cent), whereas concentrations are larger by $28\pm43$ per cent. As far as the normalisation and the slope of the $c(M,z)$ are concerned, we found $\log A = 1.62\pm 0.54$, $B=-0.88 \pm0.30$ and $C=-2.7\pm1.7$. The shifts in the central values are much smaller than the statistical uncertainties. We can conclude that the effect of priors is negligible with respect to already included statistical uncertainties.

\section{Discussion}
\label{sec_discussion}

Theoretical models of structure formation are still experiencing problems in explaining observed properties of galaxy clusters. Our study showed that the slope of the $c(M)$ relation is steep already at $z\sim 1$. 

Galaxy surveys covers a larger volume than the typical simulation box. Very massive clusters at high redshift are quite sparse in simulations and fewer than the tens we have analysed here. The two main challenges facing $N$-body analyses are the investigation of larger and larger volumes and the inclusion of baryonic physics.

The mass and concentration of clusters may be related to their recent assembly history. \citet{lud+al12} argued that the dynamical state of dark matter haloes may cause a non-monotonic relation between mass and concentration at high redshift. Massive systems at high redshift are likely caught at a transient stage of high concentration before virialization corresponding to the first pericentric passage of recently accreted material. On the other hand, dynamically relaxed systems better fits scaling laws extrapolated from less massive systems at lower redshift.

Due to their recent assembly, massive galaxy clusters at high redshift are generally expected to be out of dynamical equilibrium. haloes with masses $\gs 10^{14}M_\odot/h$ at $z\sim 1$ are very likely dynamically active. A fraction in excess of 40 per cent is observed in a state of transient maximal compression which boosts the concentration \citep{lud+al12}.

The understanding of the $c(M)$ relation is made even more complex by the fact that, for clusters past the state of maximal compression, the older the cluster, the more relaxed and more concentrated  \citep{lud+al12}. Clusters at the same concentration level might then be either very relaxed or very dynamically perturbed.

This increasingly complex picture is coherent with what found observationally in lensing studies of the mass concentration relation. Our study gave a look at what happens at redshift one. We found some suggestions that one half of the clusters is intrinsically over-concentrated. The remaining half better follows trends extrapolated from lower masses and redshifts. This might be seen as evidence that the observed sample is a mix of two populations which differ for their recent assembly history: either clusters in a transient stage of accretion or relaxed haloes. However, statistical evidence is still marginal. A better defined sample of clusters and complementary multi-wavelength observations are needed to properly assess the presence of distinct populations of clusters on an observational basis.

Baryonic physics is expected to play an important role determining the internal properties of clusters but results from numerical simulations are still ambiguous. Due to radiative cooling, baryons tend to concentrate in the inner regions, increasing the concentration of the halo. The dark matter component responds to the baryon sinking through an adiabatic contraction effect, leading to a further increase of the concentration \citep{blu+al86}. Since in smaller haloes the fraction of the mass of the central galaxy to the total mass is larger, lower mass haloes should be affected more pronouncedly. Then, the baryon effect is expected to be mass-dependent and might be at the origin of the observed steep slope in the $c(M)$ relation. 

This simplified picture does not account for feedback processes. AGN feedback, or extremely efficient feedback from massive stars reduce the baryon fraction in the inner regions of clusters. This leads to shallower inner density profiles and lower concentrations \citep{duf+al10,mea+al10}. It is then still matter of debate how cooling and feedback counterbalance to account at the same time for a steep $c(M)$ relation and the observed stellar fraction in galaxy clusters \citep{duf+al08}. Counterintuitively, \citet{deb+al13} found that the inclusion of baryon physics influences more high-mass systems than low-mass ones, due to a higher concentration of baryons in the inner regions of massive haloes, which might make the slope flatter.

\section{Conclusions}
\label{sec_conclusions}

Combined strong and weak lensing analyses make possible to accurately determine the mass and the concentration of galaxy clusters, even at very high redshift. We investigated the $c(M)$ relation in a sample of massive galaxy clusters at $z\sim 1$ by means of weak lensing and (when available) strong lensing observations. The slope of the relation is steeper than theoretical predictions, as already found for lensing samples at lower redshifts \citep{oka+al10,ogu+al12}. The staking methods do not show any over-concentration problem.

An alternative, but not mutually exclusive interpretation is that low or high concentration clusters might reflect different intrinsic properties. Dynamically active and temporarily over-concentrated clusters are more easily detected at high redshift. Clusters might be discriminated according to their dynamical status and their recent assembly history \citep{lud+al12}. If a cluster is relaxed, the more massive the less concentrated. On the other hand, clusters still accreting material may experience a transient phase of maximal compression and high concentration.

Since the paucity of multi-wavelength data, a rigorous observational classification of the dynamical state of the clusters in our sample cannot be made. We nevertheless found that clusters with prominent signs of merging activity or filamentary structure are preferentially over-concentrated.

The sample we considered is not statistical and assessing the presence of biases in the analysis is complicated by the different selection methods and observational techniques used to discover the clusters. Nevertheless, we checked for a number of systematics. Furthermore, the number of clusters considered is comparable or larger than high-redshift massive clusters usually found in numerical simulations. 

A homogeneous sample would greatly improve the statistical evidence of our analysis. Our results would significantly benefit also from a spectroscopic redshift confirmation of the detected giant arcs. This would better constrain the central slope and in turn the concentration.

\section*{Acknowledgements}
The authors thank M.~J. Jee for providing some unpublished results and L. Izzo for early discussions. This research has made use of NASA's Astrophysics Data System.


\setlength{\bibhang}{2.0em}

\end{document}